\documentclass[a4paper,11pt]{article}
\usepackage{pos}
\usepackage{slashed}

\newcommand \ml {m_l}
\newcommand \ms {m_s}
\newcommand \U {\mathcal{U}}
\newcommand \av[2] {\left\langle{#1}\right\rangle_{#2}}
\newcommand \pbp {\bar{\psi}\psi}
\newcommand \cum[1] {\mathbb{K}_{#1}}
\newcommand \lda {\lambda}
\newcommand \ru[1] {\rho_U(\lda_{#1})}
\newcommand \pu[2] {P_U(\lda_{#1};{#2})}
\newcommand \pn {P_n(\lda)}
\newcommand \fn {f_n(z)}
\newcommand \gn {g_n(\lda)}
\newcommand \nt {N_\tau}
\newcommand \ns {N_\sigma}
\newcommand \tc {T_c}
\newcommand \hlda {\hat{\lda}}
\newcommand \hp[1] {\hat{P}_{#1}(\hlda)}
\newcommand \hml {\hat{m}_l}
\newcommand \hgn {\hat{g}_n(\hlda)}
\newcommand \D {\slashed{D}}

\title{Microscopic Encoding of Macroscopic Universality in QCD Chiral Phase Transition}
\ShortTitle{Microscopic Encoding of Macroscopic Universality}

\author[1]{Heng-Tong Ding}
\author*[1]{Wei-Ping Huang}
\author[2]{Swagato Mukherjee}
\author[2]{Peter Petreczky}

\affiliation[1]{Key Laboratory of Quark \& Lepton Physics (MOE) and Institute of Particle Physics,\\
  Central China Normal University, Wuhan 430079, China}

\affiliation[2]{Physics Department, Brookhaven National Laboratory, Upton, NY 11973, USA}

\emailAdd{hengtong.ding@ccnu.edu.cn}
\emailAdd{huangweiping@mails.ccnu.edu.cn}
\emailAdd{swagato@bnl.gov}
\emailAdd{petreczk@bnl.gov}

\abstract{We reveal that the universal scaling properties of the chiral phase transition in quantum chromodynamics (QCD) at the macroscale are, in fact, encoded within the microscopic energy levels of its fundamental constituents, the quarks. We introduce a novel relation between the cumulants of the chiral order parameter, i.e., the chiral condensate, and the correlations among the energy levels of quarks, i.e., the eigenspectra of the massless QCD Dirac operator. This relation elucidates how the fluctuations of the chiral condensate arise from the correlations within the infrared part of the energy spectra of quarks, and naturally leads to a generalization of the Banks-Casher relation for the cumulants of the chiral condensate. Then, through (2+1)-flavor lattice QCD calculations using HISQ action with varying light quark masses around the chiral phase transition temperature, we demonstrate that the correlations among the infrared part of the Dirac eigenvalue spectra exhibit same universal scaling behaviors as expected of the cumulants of the chiral condensate. We find that these universal scaling behaviors extend up to the physical values of the up and down quark masses.}

\FullConference{The 40th International Symposium on Lattice Field Theory (Lattice 2023)\\
July 31st - August 4th, 2023\\
Fermi National Accelerator Laboratory\\}

\begin{document}
\maketitle

\section{Introduction}
Intensive efforts have been made to search for universal signatures of criticality in the phase diagram of strong-interaction matter governed by quantum chromodynamics (QCD), both from experimental side such as heavy-ion collisions~\cite{Luo:2017faz} and theoretically from, e.g., ab-initio lattice regularized QCD studies~\cite{Ding:2015ona}. Experimental measurements of macroscopic observables, e.g., cumulants of conserved charge distributions, did reveal certain universal behaviors in the vicinity of critical region~\cite{Luo:2017faz}; nevertheless, how such universality at macroscale arises from the fundamental constituents and interactions of QCD remains unresolved.

It is well established that for physical values of the quark masses the strongly interacting matter undergoes a rapid crossover from a hadronic phase to quark gluon plasma phase~\cite{Aoki:2006we,Bhattacharya:2014ara}; while in the limit of vanishing quark masses a true phase transition of QCD is believed to exist~\cite{Pisarski:1983ms}. In the massless limit of up and down quarks with a physical strange quark mass the QCD chiral transition is expected  to be a second order phase transition, belonging to the three-dimensional $O(4)$ universality class if axial $U_A(1)$ anomaly remains manifested around chiral transition temperature~\cite{Pisarski:1983ms,HotQCD:2019xnw,Ding:2020xlj,Ejiri:2009ac}. A peculiar feature of the second order phase transition is that around the transition region, macroscopic quantities related to the order parameter follow certain power law scaling behaviors that are uniquely characterized by the dimensionality and global symmetries of the system, irrespective of the details of its microscopic degrees of freedom and interactions. Thus lattice QCD studies of chiral phase transition using staggered fermion discretization~\cite{Kilcup:1986dg,Ejiri:2009ac,Clarke:2020htu} are expected to observe same macroscopic scaling behaviors as that, for example, in the vicinity of the liquid to superfluid $\lambda$ transition in $^4\mathrm{He}$~\cite{chaikin_lubensky_1995}, both of which belong to a 3-dimensional $O(2)$ universality class; notwithstanding, the microscopic degrees of freedom for QCD are quarks and gluons while for $^4\mathrm{He}$ are electrons and photons. Because of this universal feature, to understand and predict macroscopic properties of a system close a second order phase transition one most often resorts to a simplified effective theory possessing the same dimensionality and global symmetries of the original theory, ignoring its microscopic complexities~\cite{chaikin_lubensky_1995}. However, it is still unclear if and how the macroscopic universal scaling properties of the strong interaction in the vicinity of the chiral phase transition are concealed within the building blocks of QCD.

Some light on this question is shed by the Banks-Casher relation~\cite{Banks:1979yr} that relates the macroscopic chiral order parameter, the chiral condensate, to the microscopic zero mode of Dirac eigenspectra in the chiral limit. Inspired by this, the goal of this work is lattice QCD-based understanding of possible connections between the universal features at macroscale and microscopic scales of QCD, i.e., the energy spectrum of quarks. An analogous goal for quantum electrodynamics will be to comprehend how the macroscopic scaling properties near the $\lambda$ transition of $^4\mathrm{He}$ arise from the energy levels of electrons without resorting to an effective theory.

This proceeding is organized as follows. In section 2 theoretical relations between cumulants of chiral condensate and correlations among eigenspectra of massless Dirac operator are established. Section 3 shows lattice setup. Section 4 demonstrates how the $O(2)$ scaling properties of the cumulants of chiral condensate are reflected within the correlations of Dirac eigenspectra through the state-of-the-art lattice QCD calculations in the staggered discretization scheme. Finally conclusions are given in Section 5. The detailed information about this work can be found in Ref.~\cite{Ding:2023oxy}.

\section{Theoretical developments}
For (2+1)-flavor QCD with degenerate light up ($u$) and down ($d$) quarks having masses $\ml=m_u=m_d$ and a heavier strange quark with physical mass $\ms$, the strange part has negligible impact on the discussion of the $O(2)$ critical behaviors and thus the following main theoretical idea is developed by just considering QCD with two degenerate light quarks.

To probe a system in the chiral limit $\ml \to 0$, a probe operator $\pbp(\epsilon)\equiv 2\text{Tr}(\D[\U] + \epsilon)^{-1}$ is introduced, where the valance quark mass, $\epsilon>0$, is used to facilitate the evaluation of the operator. Here $\U$ is a background $SU(3)$ gauge field distributed according to $\exp\{-S[\U,0]\}$ with $S[\U,\ml] = S_g[\U] + \bar{\psi} \D[\U]{\psi} + \ml\pbp$ the Euclidean QCD action, where $\pbp = \bar\psi_u\psi_u + \bar\psi_d\psi_d$ and $S_g[\U]$ is the pure gauge action; $\D[\U]$ is the massless QCD Dirac operator for a given $\U$; $\mathrm{Tr}$ denotes traces over the color, spin and space-time indices.

By defining a generating functional expressed as
\begin{align}
  \mathbb{G}(\ml;\epsilon) = \ln \av{ \exp \left\{ - \ml \pbp(\epsilon) \right\} } {0} \,,
\label{eq:gf1}
\end{align}
the $n^\text{th}$ order cumulant, $\cum{n}$, of the order parameter $\pbp(\ml)$ for the chiral phase transition can be obtained
\begin{align}
  \cum{n} \left[ \pbp \right]= \frac{T}{V} \, (-1)^n \left. \frac{\partial^n \mathbb{G}(\ml;\epsilon)}{\partial \ml^n} \right\vert_{\epsilon=\ml} \,.
\label{eq:kn1}
\end{align}
Here $T$ is the temperature and $V$ is the spatial volume of the system, and $\av{\cdot}{0}$ denotes expectation value with respect to the QCD partition function in the chiral limit, $Z(0)=\int \exp\{-S[\U,0]\} \mathcal{D}[\U]$. With $\av{\cdot}{}$ the expectation value with respect to the QCD partition function $Z(\ml)=\int \exp\{-S[\U,\ml]\} \mathcal{D}[\U]$ and recognizing $\av{\mathcal{O}}{} = \av{ \mathcal{O} \exp\{ - \ml \pbp(\ml) \} } {0}/\av{\exp\{ - \ml \pbp(\ml) \} } {0}$ and consequent $Z(\ml)/Z(0)= \av{ \exp\{ - \ml \pbp(\ml) \} } {0}$, it is evident that $\cum{n}$ are the standard cumulants of $\pbp(\ml)$; e.g., $\cum{1}\left[ \pbp \right]=T\langle\pbp(\ml)\rangle/V$, $\cum{2}\left[ \pbp \right]=T\langle[\pbp(\ml)-\langle\pbp(\ml)\rangle]^2\rangle/V$, $\cum{3}\left[ \pbp \right]=T\langle[\pbp(\ml)-\langle\pbp(\ml)\rangle]^3\rangle/V$.

Energy levels of a massless quark in the background of $\U$ are given by the eigenvalues, $\lda_j[\U]$, of $\D[\U]$. In terms of $\lda_j[\U]$ the probe operator can be formulated as $\pbp(\epsilon)\equiv 2\text{Tr}(\D[\U] + \epsilon)^{-1} = 2 \sum_j  (i\lda_j+\epsilon)^{-1}$. Thus, \autoref{eq:gf1} becomes
\begin{align}
  \mathbb{G}(\ml;\epsilon) = \ln \av{ \exp \left\{ - \ml \int_0^\infty \negthickspace \negthickspace \pu{}{\epsilon} d\lda  \right\} } {0} \,,
\label{eq:gf2}
\end{align}
where 
\begin{align}
  \pu{}{\epsilon} = \frac{4\epsilon\ru{}}{\lda^2 + \epsilon^2} \,, 
  ~ \text{and} ~
  \ru{} = \sum_j \delta(\lda-\lda_j) \,.  
\label{eq:pu}
\end{align}
From \autoref{eq:kn1} it is straightforward to obtain
\begin{align}
  \cum{n} [\pbp] = \int_0^\infty \negthickspace \negthickspace \pn d\lda  \,,
\label{eq:kn2}
\end{align}
where $P_1(\lda)=K_1[\pu{}{\ml}]$ for $n=1$, and for $n\geq2$
\begin{align}
  P_n(\lda)=\int_0^\infty & K_1\big[\pu{}{\ml}, \pu{2}{\ml}, \dotsc, \pu{n}{\ml} \big] \prod_{i=2}^{n}d\lda_i  \,.
\label{eq:pn}
\end{align}
Here $K_1$ denotes the 1$^\text{st}$ order joint cumulant of $n$ variables ($X_i$) defined as 
\begin{align}
  K_1 (X_1,\cdots,X_n)= \frac{T}{V} \, (-1)^n \left. \frac{\partial^n \ln\av{\prod_{i=1}^{n} e^{-t_i X_i}}{} }{\partial t_1 \dotsm \partial t_n}  \right\vert_{t_1, \dotsc,t_n=0} \,.
\label{eq:K1}
\end{align}

\autoref{eq:kn2} is our main theoretical result connecting the cumulants of the order parameter to the $n$-point correlations of the quark energy levels $\ru{}$. 
The cumulants of the light quark chiral condensate $\cum{n}[\pbp]$ are explicitly expressed in terms of $\ru{}$, e.g., for $n \leq$ 3, as follows:
\begin{align}
\begin{split}
   \cum{1}[\pbp] &
   = \frac{T}{V}\int_0^\infty \negthickspace \negthickspace \mathrm{d} \lda \frac { 4\ml \av{\ru{}}{} } {\lda^2+\ml^2} \,, 
\\
   \cum{2}[\pbp] &
   = \frac{T}{V} \int_0^\infty \negthickspace \negthickspace \mathrm{d}\lda_1 \mathrm{d}\lda_2 \frac{(4\ml)^2} {(\lda_1^2+\ml^2)(\lda_2^2+\ml^2)} \left[ \av{\ru{1} \ru{2}}{} - \av{\ru{1}}{} \av{\ru{2}}{} \right] \,, 
\\
   \cum{3}[\pbp] &
   = \frac{T}{V} \int_0^\infty \negthickspace \negthickspace \mathrm{d}\lda_1 \mathrm{d}\lda_2 \mathrm{d}\lda_3 \frac{(4\ml)^3} {(\lda_1^2+\ml^2)(\lda_2^2+\ml^2)(\lda_3^2+\ml^2)} \bigg[\av{\ru{1} \ru{2} \ru{3}}{} 
   \\ & - \av{\ru{1} \ru{2}}{}\av{\ru{3}}{} - \av{\ru{1} \ru{3}}{}\av{\ru{2}}{} 
   \\ & - \av{\ru{2} \ru{3}}{}\av{\ru{1}}{} + 2\av{\ru{1}}{}\av{\ru{2}}{}\av{\ru{3}}{}\bigg] \,.
\end{split}
\label{eq:eq:Kn-rhoU}
\end{align}

The chiral phase transition in the staggered lattice QCD at nonvanishing lattice spacings is expected to be in the 3-dimensional $O(2)$ universality class. Following the expectations for a 3-dimensional $O(2)$ spin model in the vicinity of the phase transition~\cite{Engels:2001bq} the $n^{\rm th}$ order cumulant of chiral condensate can be expressed as follows
\begin{align}
  \cum{n}[\pbp] = \int_0^\infty \negthickspace \negthickspace \pn d\lda \sim \ml^{1/\delta-n+1} f_n(z) \,.
\label{eq:scaling}
\end{align}
Here the scaling variable $z \propto z_0 \ml^{-1/\beta\delta}(T-\tc)/\tc$, where $\tc$ is the chiral phase transition temperature and $z_0$ is a scale parameter; both are non-universal but system (QCD) specific. $\beta$ and $\delta$ are the universal critical exponents, and $f_{n+1}(z) = (1/\delta-n+1) f_{n}(z) - z f_{n}^{\prime}(z) / {\beta\delta}$ are the universal scaling functions of 3-dimensional $O(2)$ universality class. Here, $n\ge1$ and the superscript prime denotes derivative with respective to $z$~\footnote{According to the notation in the literature $f_1(z)\equiv f_G(z)$ and $f_2(z)\equiv f_\chi(z)$.}. In our work we adopted $\beta=0.349$, $\delta=4.78$ and for consistency the scaling functions $f_n(z)$ of the $O(2)$ universality class determined from Refs.~\cite{Engels:2001bq,Ejiri:2009ac}. 

\autoref{eq:scaling} indicates that universal scaling properties of the macroscopic observables, $\cum{n}[\pbp]$, arise from the correlations among the microscopic energy levels, $\pn$. To elucidate this point, we consider the case as $\ml\to0$. From \autoref{eq:pu} one finds $\pu{}{\epsilon\to0} = 2\pi \ru{} \delta(\lda)$, giving $\lim_{\ml\to0}P_1(\lambda)=2\pi K_1[\ru{}]\delta(\lda)$ and $\lim_{\ml\to0} \pn = (2\pi)^n K_1[\ru{}, (\rho_U(0))^{n-1}] \delta(\lda)$ for $n\geq2$ (from \autoref{eq:pn}). Noting that $K_1$ of $n$ identical variables is equivalent to $\cum{n}$, in the chiral limit \autoref{eq:kn2} thus becomes a generalization of the Banks-Casher relation~\cite{Banks:1979yr} expressed as follows
\begin{align}
  \lim_{\ml\to0} \cum{n} [\pbp] = (2\pi)^n \cum{n} [\rho_U (0)] \,.
\label{eq:genBC}
\end{align}
To the best of our knowledge this generalized relation between the higher order cumulants of chiral condensate in the chiral limit and density of the deep infrared energies of quarks is new in literature. 

In the chiral limit  and close to $\tc$, $\cum{n}[\pbp]$ should manifest universal scaling, e.g., $\cum{1}[\pbp] \sim |{(T-\tc)}/{\tc}|^{\beta}$ and $\cum{2}[\pbp] \sim |{(T-\tc)}/{\tc}|^{\beta(1-\delta)}$. According to \autoref{eq:genBC} this must arise from the universal behaviors of the $\lambda$-independent $\cum{n}[\rho_U(0)]$. Thus, it is natural to expect for small $\ml$ within the scaling window the critical scaling of $\cum{n}[\pbp]$ in \autoref{eq:scaling} arises from the universal behaviors of the amplitudes of $\pn$ at the infrared, and not from its system-specific $\lambda$ dependence; i.e., $\pn = \ml^{1/\delta-n+1} \fn \gn$, where $\gn$ are nonuniversal functions encoding the properties of the specific system under consideration.

\section{Lattice setup}
To numerically establish our conjecture for $\pn$, lattice QCD calculations were carried out between $T=135-176$~MeV for (2+1)-flavor QCD using HISQ action~\cite{Ding:2023oxy}. The $\ms$ was fixed to its physical value with a varying $\ml = \ms/27, \ms/40, \ms/80, \ms/160$, which correspond to the Goldstone pion masses $m_{\pi}\approx$ 140, 110, 80, 55~MeV, respectively. The temporal extent of the lattices were $\nt=8$, and spatial extents were chosen to be $\ns=(4-7)\nt$. Observables were calculated on gauge configurations from every 10$^\text{th}$ molecular dynamics trajectory of unit length, after skipping at least first $800$ trajectories for thermalization. $\ru{}$ and $\pn$ for $n = 1, 2, 3$ over the entire range of $\lda$ were computed using the Chebyshev filtering technique combined with the stochastic estimate method~\cite{Ding:2020eql, Ding:2020xlj, Ding:2021gdy, Giusti:2008vb, Cossu:2016eqs, Fodor:2016hke} on about $3000$ configurations. Orders of the Chebyshev polynomials were set to $2\times10^5$ and $24-96$ Gaussian stochastic sources were used. $\hat{\mathbb{K}}_n$ for $n\leq 3$ were also directly computed via the inversion of the light fermion matrix on each of these ensembles using 50 Gaussian random sources.

\section{Results}
Owing to \autoref{eq:genBC} we expect the relevant infrared energy scale is $\lda\sim\ml$ for small values of $\ml$. It is thus natural to express all quantities as functions of the dimensionless and renormalization group invariant $\lda/\ml$, and following notations are used hereafter: 
\begin{align}
\begin{split}
  & \hlda = \lda / \ml \,, \quad 
  \hml = \ml / \ms \,, z=z_0\hml^{-1/\beta\delta}(T-\tc)/\tc\,,\\
  & \hp{n} = \ms^{n+1} \hml P_n(\lda) / \tc^4 \,, 
  \quad \text{and} \, \quad 
  \mathbb{\hat K}_{n} [\pbp] = \ms^n \cum{n} [\pbp] / \tc^4 = \int_0^\infty \negthickspace \negthickspace \hp{n} d\hlda \,.
\end{split}
\label{eq:hatdef}
\end{align}
The system-specific parameters $\tc=144.2(6)$ MeV and $z_0=1.83(9)$ needed below to obtain $f_n(z)$ were taken from Ref.~\cite{Clarke:2020htu}, where 3-dimensional $O(2)$ scaling fits were carried out for the same lattice ensembles but using an entirely different macroscopic observable, namely the $\ml$ dependence of the static quark free energy.

Due to the explicit quark mass dependence of the order parameter in the chiral phase transition, the $n^{\text{th}}$ order derivative of chiral condensate with respect to quark mass (i.e., the symmetry breaking field) always consists of two contributions: one of which is a disconnected part, i.e., $\cum{n} \left[ \pbp \right]$ defined in \autoref{eq:kn1}, corresponding to the fluctuation of quark condensate and has a direct counterpart in a spin model; the other is connected part specific to QCD. Thus strictly speaking the right hand side of \autoref{eq:scaling}, i.e., $\hml^{1/\delta-n+1} f_n(z)$, only describes the scaling of the full derivatives of chiral condensate containing both disconnected and connected parts in the vicinity of the chiral transition. 

\begin{figure}[!htp]
  \centering
    \includegraphics[width=0.32\textwidth,height=0.15\textheight]{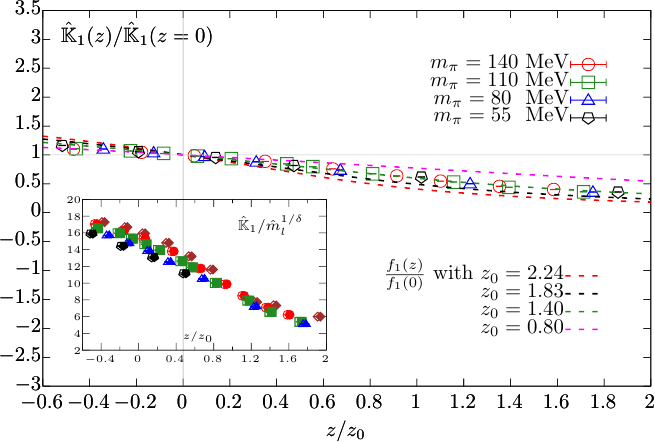}
    \includegraphics[width=0.32\textwidth,height=0.15\textheight]{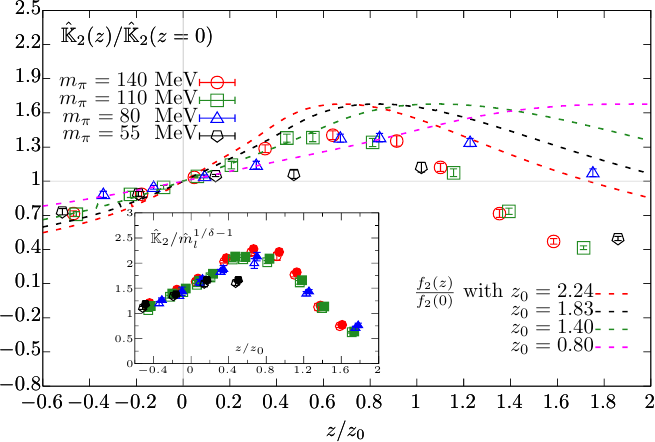}
    \includegraphics[width=0.32\textwidth,height=0.15\textheight]{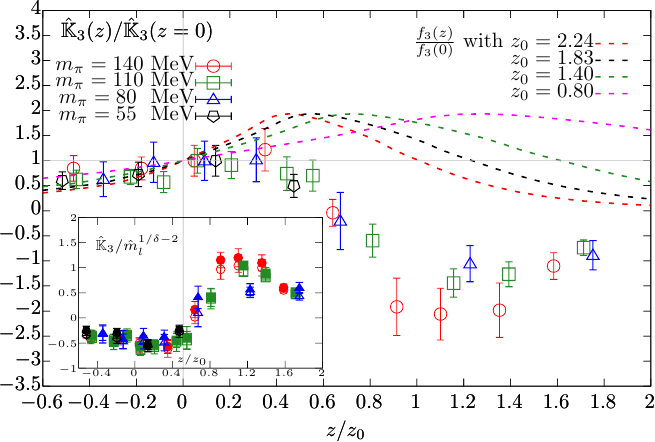}
  	\caption{Comparisons of ratio $\hat{\mathbb{K}}_n(z)/\hat{\mathbb{K}}_n(z=0)$ as a function of $z/z_0$ and corresponding ratio of scaling function $f_n(z)/f_n(z=0)$ for $n=1$ (left), 2 (middle) and 3 (right) with different values of $z_0$. The inserts show $\hat{\mathbb{K}}_n$ normalized by $\hat{m}_l^{1/\delta-n+1}$, where filled points (slightly shifted horizontally for visibility) denote $\hat{\mathbb{K}}_n$ obtained from $\rho_U(\lambda)$ using \autoref{eq:eq:Kn-rhoU} while open points are those computed via the inversion of fermion matrix.}
	\label{fig:K123-reproduce}
\end{figure}

To test if $\mathbb{\hat K}_{n}$ for $n\geq2$ as merely a disconnected part can still be described by aforementioned scaling, \autoref{fig:K123-reproduce} shows the ratio of $\mathbb{\hat K}_{n}$ to its value at $\tc$ as a function $z/z_0$. It can be observed that for $|z/z_0|\lesssim0.2$, $\mathbb{\hat K}_{n}(z)/\mathbb{\hat K}_{n}(z=0)$ obtained from pion masses ranging from 140 MeV to 55 MeV overlap with each other and can be well described by the curves of $f_n(z)/f_n(z=0)$ irrespective of various values of $z_0$ in the range of $[0.8,2.24]$. In the inserts $\mathbb{\hat K}_{n}$ rescaled by $\hat{m}_l^{1/\delta-n+1}$ shows again the quark mass independence in the range of $|z/z_0|\lesssim0.2$, except for chiral condensate $\mathbb{\hat K}_{1}$ due to remanent mass contribution from its ultraviolet part. These two observations suggest that $\mathbb{\hat K}_{n} \propto \hml^{1/\delta-n+1} f_n(z)$ within the critical region $|z/z_0|\lesssim0.2$. 

The inserts of \autoref{fig:K123-reproduce} also show that integrals over the relevant nonvanishing infrared regions of $\hp{n}$ (cf. \autoref{eq:kn2}) reproduce $\mathbb{\hat K}_{n}$, independently calculated through inversions of the fermion matrices, for $n\leq 3$. The consistency of $\mathbb{\hat K}_{n}$ given by the two different methods verifies the reliability of the numerical results of $\hp{n}$ with $n=1,2$ and 3.

\begin{figure*}[!htp]
  \centering
    \includegraphics[width=0.32\textwidth,height=0.15\textheight]{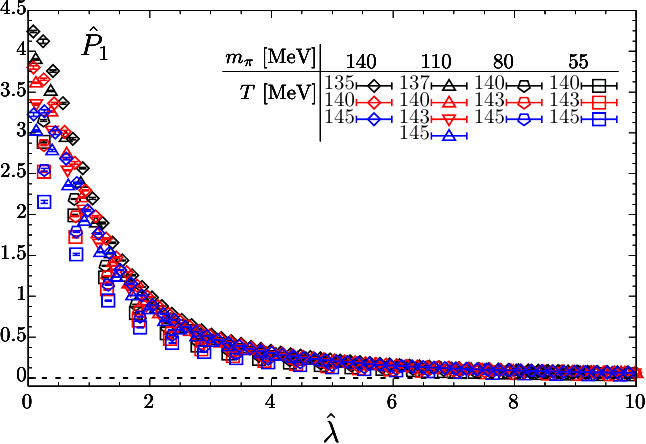}
    \includegraphics[width=0.32\textwidth,height=0.15\textheight]{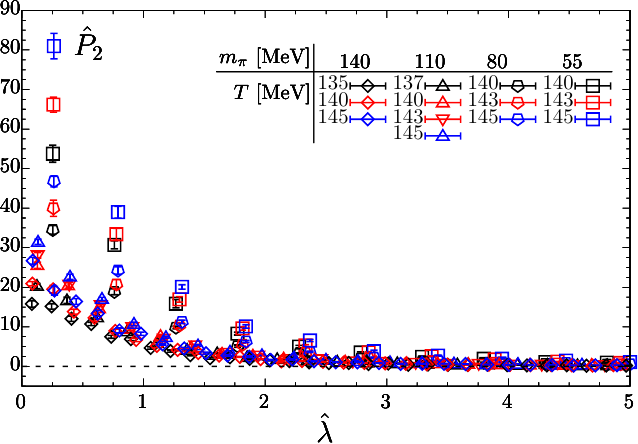}
    \includegraphics[width=0.32\textwidth,height=0.15\textheight]{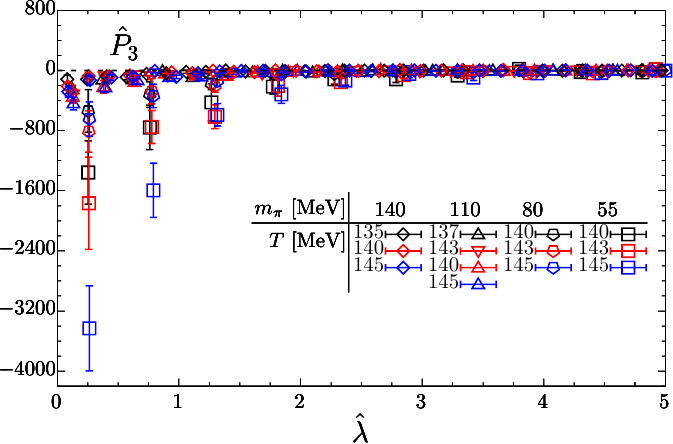}
  	\caption{$\hp{1}$ (left), $\hp{2}$ (middle) and $\hp{3}$ (right) for 135~MeV~$\le T \le$~145~MeV and 55~MeV~$\le m_\pi \le$~ 140~MeV.}
	\label{fig:P123}
\end{figure*}

In \autoref{fig:P123} we show $\hp{n}$ for $n\leq 3$ as a function of $\hlda$ in the proximity of $\tc$. $\hp{n}$ rapidly vanishes for $\hlda\gtrsim1$, and the regions where $\hp{n}\ne0$ get smaller with increasing $n$. This reinforces that the relevant infrared energy scale turns out to be $\hlda\sim1$. In this infrared region $\hp{n}$ at a fixed $T$ shows clear dependence on $\ml$, which becomes stronger for increasing $n$. The form of $\ml$ dependence of $\hp{n}$ also changes with varying $T$. As expected our results become increasingly noisy with increasing $n$ and decreasing $\ml$. With our present statistics we cannot access correlation functions $n>3$, particularly for smaller $\ml$.

\begin{figure*}[!htp]
  \centering
    \includegraphics[width=0.32\textwidth,height=0.15\textheight]{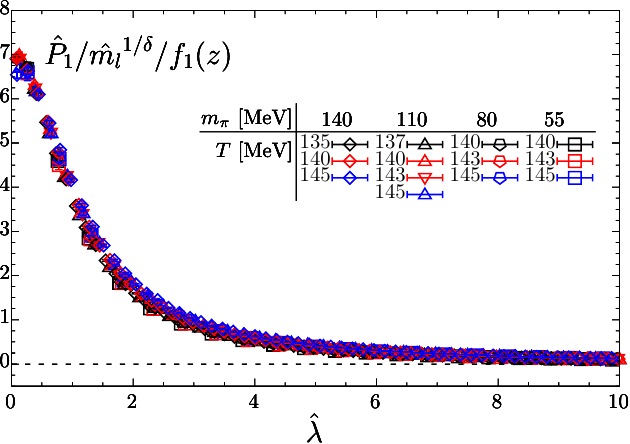}
    \includegraphics[width=0.32\textwidth,height=0.15\textheight]{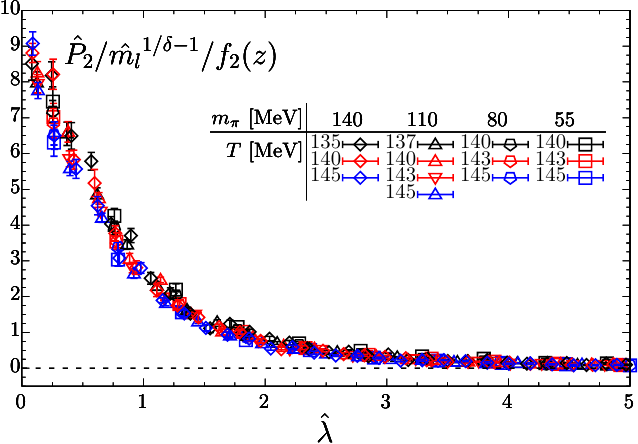}
    \includegraphics[width=0.32\textwidth,height=0.15\textheight]{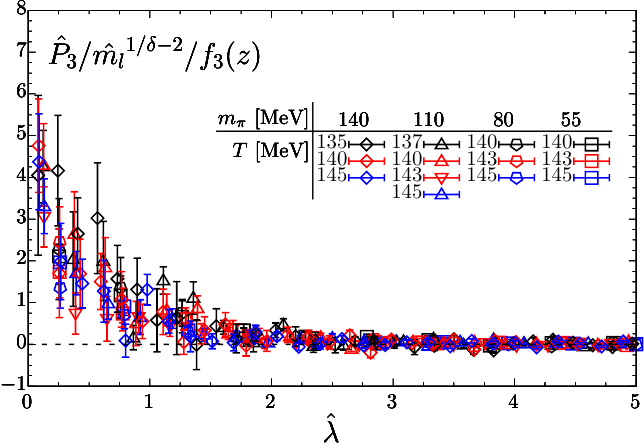}
  	\caption{$\hp{n}$ in \autoref{fig:P123} rescaled by $\hml^{1/\delta+1-n}f_n(z)$ for $n=1$ (left), $n=2$ (middle) and $n=3$ (right).}
 \label{fig:RescaledP123}
\end{figure*}

The $\ml$ and $T$ dependence of $\hp{n}$ shown in \autoref{fig:P123} can be understood in terms of the 3-dimensional $O(2)$ scaling properties. Once the $\hp{n}$ are rescaled with respective $\hml^{1/\delta+1-n}f_n(z)$ the data in \autoref{fig:P123} magically collapse onto each other, see \autoref{fig:RescaledP123}. Thus, our expectations for $\hp{n}$ are clearly borne out in \autoref{fig:RescaledP123}, namely 
\begin{align}
  \hp{n} = \hml^{1/\delta-n+1} \fn \hgn \,,
\label{eq:pn2}
\end{align}
where $\hgn$ characterize the system specific of the $n$th order energy-level correlations. To satisfy our generalized Banks-Casher relations of \autoref{eq:genBC} the $\hgn$ must also satisfy $\lim_{V\to\infty} \lim_{a\to0} \lim_{\ml\to0} $ ~ $ \hgn \to \delta(\hlda)$, such that $\cum{n}[\pbp]$ has the correct scaling behavior in $(T-\tc)/\tc$.

The values of $z_0$ and $\tc$ used to demonstrate the universal scaling in \autoref{fig:RescaledP123} were obtained by fitting lattice results for $\ml$ dependence of the static quark free energy only for 55~MeV~$\le m_\pi \le$~110~MeV~\cite{Clarke:2020htu}. It is noteworthy that the physical QCD with $m_\pi\approx$~140~MeV also shows the same universal scaling for 135~MeV~$\le T \le$~145~MeV. Outside this temperature window we do not observe scaling as shown in the Supplemental Material of Ref.~\cite{Ding:2023oxy}.

As mentioned in Ref.~\cite{Clarke:2020htu}, $\{\tc,z_0\}$ are not very well determined at present. By using other values for $\{\tc,z_0\}$ quoted in Ref.~\cite{Clarke:2020htu} we checked that the scaling of \autoref{fig:RescaledP123} is fairly insensitive to the exact values of $\{\tc,z_0\}$ as shown in the Supplemental Material of Ref.~\cite{Ding:2023oxy}. Presumably, this is because $\hp{n}$ are sensitive only to the deep infrared physics $\lda\sim\ml$. This is in contrast to many other macroscopic quantities used for detailed scaling studies, that contain large contributions from the ultraviolet energy scale~\cite{HotQCD:2019xnw,Ejiri:2009ac,Hegde:2015tbn,Dini:2021hug,Kaczmarek:2021ufg,Kotov:2021rah}. This suggests that it even might be advantageous to use $\hp{n}$ for future detailed scaling studies to determine the QCD specific parameters, such as $T_c$
and $z_0$.

\section{Conclusions}
In this work we investigate how the universal critical scaling of macroscopic observables near the QCD chiral transition arises from the microscopic degrees of freedom. We have presented a theoretical connection between the $n^\text{th}$ order cumulant of the chiral order parameter and the $n$-point correlations of the quark energy spectra. This connection led us to a generalized Banks-Casher relation, equating the $n^\text{th}$ order cumulant of chiral condensate to the $n^\text{th}$ order cumulant of the zero mode of the quark energy in the chiral limit. These new theoretical developments establish a direct connection between the universal scaling observed at the macroscale and the microscopic energy levels of the system. Through staggered lattice QCD calculations in the vicinity of the chiral phase transition with a series of light quark masses we have discovered the hidden universality within the correlations among the quark energy spectra. We have found that these universal behaviors are also imprinted within the microscopic energy levels of QCD with physical light quark masses. 

\acknowledgments
We thank Yu Zhang for early involvement, Sheng-Tai Li for technical support, Jacobus Verbaarschot and members of HotQCD collaboration for discussions. 
This work is supported partly by the National Key Research and Development Program of China under Contract No. 2022YFA1604900; the National Natural Science Foundation of China under Grants No.~12293064, No.~12293060 and No.~12325508, and by the U.S. Department of Energy, Office of Science, Office of Nuclear Physics through Contract No.~DE-SC0012704 and within the framework of Scientific Discovery through Advance Computing (SciDAC) award Fundamental Nuclear Physics at the Exascale and Beyond. 
Numerical simulations were performed on the GPU cluster in the Nuclear Science Computing Center at CCNU (NSC$^3$), Wuhan supercomputing center, the facilities of the USQCD Collaboration funded by the Office of Science of the U.S. Department of Energy.

\bibliographystyle{JHEP.bst}
\bibliography{ref.bib}

\end{document}